\documentclass[useAMS,usenatbib,usegraphicx,epsfig]{mn2e}

\usepackage{amssymb}
\usepackage{graphicx}
\usepackage{amsmath}
\usepackage{mathrsfs}
\usepackage{amsmath}

\title[$z$ Filter and Photo-z]{Improved Photometric Redshifts via Enhanced Estimates of System Response, Galaxy Templates, and Magnitude Priors}

\author[Schmidt \& Thorman]{Samuel J.~Schmidt\thanks{Email:sschmidt@physics.ucdavis.edu} and  
Paul Thorman\thanks{Email:thorman@physics.ucdavis.edu}\\
\noindent Dept. of Physics, University of California, One Shields Ave., Davis, CA 95616}

\begin{document}

\maketitle

\begin{abstract}
Wide, deep photometric surveys require robust photometric redshift estimates (photo-z's) for studies of large-scale structure.  These estimates depend critically on accurate photometry. We describe the improvements to the photometric calibration and the photo-z estimates in the Deep Lens Survey (DLS) from correcting three of the inputs to the photo-z calculation: the system response as a function of wavelength, the spectral energy distribution templates, and template prior probabilities as a function of magnitude.  We model the system response with a physical model of the MOSAIC camera's CCD, which corrects a 0.1 magnitude discrepancy in the colours of type M2 and later stars relative to the SDSS $z$ photometry. We provide our estimated $z$ response function for the use of other surveys that used MOSAIC before its recent detector upgrade.  The improved throughput curve, template set, and Bayesian prior lead to a 20 per cent reduction in photo-z scatter and a reduction of the bias by a factor of more than two.  This paper serves as both a photo-z data release description for DLS and a guide for testing the quality of photometry and resulting photo-z's generally.
\end{abstract}


\section{Introduction}
Due to the time and expense of obtaining spectra of large numbers of faint galaxies, deep photometric surveys have increasingly turned to photometric redshift estimates (photo-z's). For the next generation of all-sky deep surveys, such as LSST \citep{lsst}, Euclid \citep[]{Lau:10}, the Dark Energy Survey \citep{DES}, and Pan-STARRS \citep{panstarrs}, photometric redshifts will be essential to the study of large-scale structure, baryon acoustic oscillations, and the evolution of the Universe over time.  The improved photo-z's described in this paper, which are based on photometry from the Deep Lens Survey \citep[DLS;][]{wittman02}, have enabled the first tomographic measurement of weak lensing magnification \citep[]{Mor:12}, made it possible to examine galaxy-mass correlations at large scales \citep[]{Choi:12}, and been used to establish joint constraints on $\Omega_{M}$ and $\sigma_{8}$ using cosmic shear \citep[]{Jee:12}.

Fast and robust photo-z software \citep[e.g., BPZ, Hyper-z;][]{BPZ:00, Hyperz:00} can generate detailed redshift estimates based on measured galaxy fluxes in different passbands. A popular method uses spectroscopic templates based on nearby galaxies \citep[e.g.,][hereafter CWW]{CWW:80}, redshifted and convolved with user-supplied system response functions to generate theoretical colours for each galaxy spectral type as a function of redshift. These are often combined with a Bayesian prior based on the magnitudes (apparent or absolute) and relative abundances of the various spectral types to determine  the relative posterior probability of each type and redshift for a given galaxy's measured fluxes and flux errors.

Good photometry is essential for accurate photo-z's.  In any photo-z procedure that fits spectral energy distributions (SEDs), the key assumption is that the input fluxes are sampling light from a single isolated galaxy. Any flux missing in a particular band from improper deblending, seeing differences, or other unrecognized errors will alter the galaxy's colour, and will very likely translate into an error in the inferred redshift.  This was specifically demonstrated for various methods of compensating for seeing in \citet{hildebrandt12}. In addition, errors in the determination of the system response functions (i.e.,~filter+mirror+CCD~response+atmosphere) will propagate into the photo-z calculation through the model flux calculations.  

These errors are amplified when ``training set" data from the survey itself are used to optimize photo-z performance, which is done in this paper and elsewhere \citep[]{Bud:00,Ilb:06}.  Systematic biases present in the photometry can propagate through the analysis and degrade the photo-z performance.  
In this paper we will discuss one such example: the incorrect specification of the effective system response for the DLS $z$-band.  After correcting the system response curve, we then discuss modifications of the photo-z template set and Bayesian prior for the DLS. 

Details of the DLS photometric calibration can be found in the DLS ubercal paper \citep{wittman11} and the DLS data release paper (Wittman et al.~in prep.).  The effects of the prior probability and the error distribution have previously been studied for the Deep Lens Survey data \citep{witt07,marg08}.  In this paper, in addition to constructing a new prior, we discuss the diagnosis of photometry problems and a technique to correct the filter response curve (\S\ \ref{through}).  In \S\,\ref{tweak} we discuss modifications to SED templates for the photo-z calculation, in \S\,\ref{prior} we construct a prior based on multiple datasets, and in \S\,\ref{results} we describe the resulting DLS photo-z's.  We conclude and discuss future work in \S\,\ref{conclusions}.

\section{Summary of Observations}

\subsection{Imaging}

Our analysis is focused on the Deep Lens Survey, a deep $BVRz$ imaging survey carried out with the MOSAIC and MOSAIC-II cameras at KPNO and CTIO between 1999 and 2006. The primary imaging covered 20 deg$^2$ in five 2$^{\circ}$ x 2$^{\circ}$ fields spaced around the sky. The fields were chosen for their lack of bright stars and galaxies, their appropriate position for split-night observations, and the presence of pre-existing deep spectroscopic surveys, when possible. The $R$-band images were taken during periods of good seeing (FWHM$<0.9''$) with a total integration time of 24~ks, while the $B$, $V$, and $z$ bands were integrated for 18~ks apiece. Average 5$\sigma$ detection limits for point sources in $B$, $V$, $R$, and $z$ are 26.5, 26.5, 27, and 24 magnitudes, respectively (all magnitudes on the Vega system), with some minor variations in depth due to variable seeing.  There are also shallower areas around the edges of the dithered pointings (``subfields"), which received fewer observations than the subfield centres. Magnitudes and colours were measured using Colorpro \citep{colorpro}, which corrects for differences in PSF between filters as well as differences in seeing between stacks of images taken at different times. These magnitudes were used to generate improved flat field maps for the MOSAIC instruments \citep[][]{wittman11} using the ubercal method of \citet{ubercal}. The complete reduction pipeline and cataloguing procedure will be explained in a companion paper to the DLS public data release (Wittman et al.~in prep.). 

\subsection{Spectroscopy}\label{spectro}

In addition to measured fluxes, the template adjustment method requires that the redshifts of a fiducial sample be known. 
The DLS Field 5 (F5) imaging was used to select targets for a subset of the PRIsm MUlti-object Survey \citep[PRIMUS,][]{coil:11}.  PRIMUS is a low-resolution objective prism slit mask spectroscopic survey using the IMACS instrument on the 6.5-metre Magellan telescope.  The instrument has spectroscopic resolution R$\sim 40$, resulting in redshifts precise to $\sigma_{z}=0.005/(1+z)$.   For F5, PRIMUS is complete to $R=22.8$ and observed a random sample consisting of 30 per cent of objects  with $24.0 > R > 22.8$.   We begin with 10,695 objects in F5, and after conservative masking and a cut on quality flag for reliable redshifts ($\textrm{ZCONF}>2$) we are left with a sample of 9107 galaxies, 703 of which are at $R>23.3$.

In DLS Field 2 (F2), we made use of targeted redshift observations from the Smithsonian Hectospec Lensing Survey \citep[SHELS,][]{Gel:05,Gel:10}, which used Hectospec, a moderate-resolution fiber-fed spectrograph on the 6.5-metre MMT, to measure redshifts for galaxies with $R<20.3$ selected from the DLS R-band imaging. Based on repeated galaxy observations, they estimate their redshift error (stated as $cz(1+z)^{-1}$) to be 27 km s$^{-1}$ for emission line objects and 37 km s$^{-1}$ for absorption line objects, negligible for our purposes.

\begin{figure*}
 \centering
\includegraphics[width=.99\hsize]{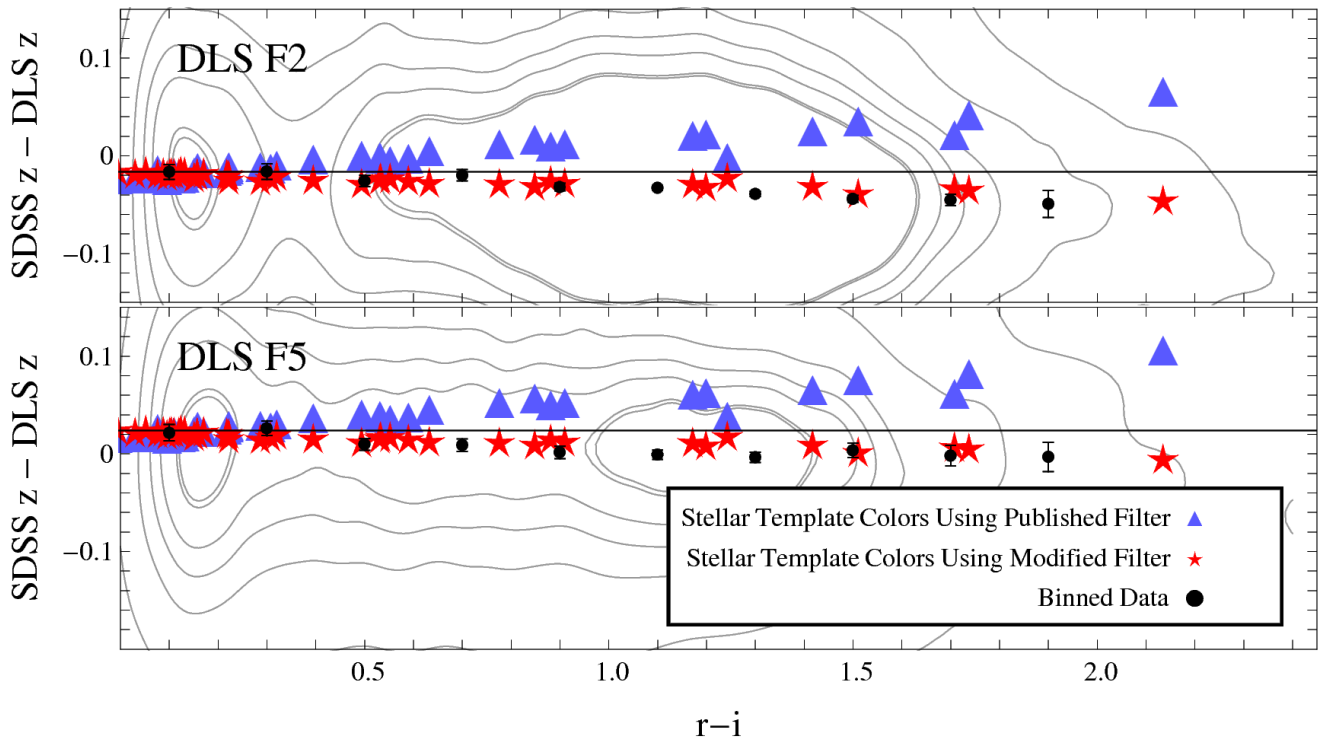}
\caption{SDSS $z$--DLS $z$ as a function of SDSS $r-i$ colour. Contours show a kernel density estimator for the colours of the stars, with contours at 0.99, 0.95, 0.68, 0.5, 0.32, 0.05, and 0.01 of the total probability. Mean stellar colours in bins are plotted as black points with error bars. Blue triangles mark the theoretical colours of stars using the published version of the system response function; red stars show the colours with our modified response. F2 and F5 were imaged from KPNO and CTIO respectively, with different MOSAIC cameras, but the modified response function better predicts our measured stellar colours in both cases.  \label{starfig}}
\end{figure*}

\section{Correcting the Throughput Curve}
\label{through}
\subsection{Identifying problems with photometry}
The DLS was originally calibrated, in part, by comparing the colours of pointlike sources measured in the science images to the predicted colours of stars generated by integrating library stellar spectra with the system throughput as a function of wavelength. This measured ``stellar locus'' was used to correct colour measurements across different fields and subfields of the DLS. 

In order to compare the DLS photometry to an independent standard, we make use of the eighth release of the Sloan Digital Sky Survey photometry \citep[SDSS;][]{aihara11}; however, the two surveys use different filters and CCDs, so the magnitudes of stars in one survey must be converted to equivalent magnitudes in the other before a proper comparison can be made. The magnitude difference depends both on the response of each system and on the SED of the target, so we focus on the stars of our own Galaxy as targets of approximately known SED. The SDSS colours of various spectral types of star have been calculated and measured directly \citep{covey07}, and the $(r-i)$ colour was shown to be a monotonic function of spectral type.

Using $(r-i)$ as a proxy for spectral type, we plot the difference between our DLS $z$ magnitudes and the SDSS $z$ magnitudes for well-detected stars (S/N $>10$ in SDSS $r$, $i$, and $z$) as a function of $(r-i)$ in Figure~\ref{starfig}. The upper and lower panels in Figure~\ref{starfig} are star samples from fields F2 and F5, which were observed from KPNO and CTIO respectively, so the similarity of the colours indicates that the two MOSAIC cameras had similar reseponse functions at the time of the survey. After spatially matching SDSS objects with high S/N to objects in the DLS catalogue in the area of overlap (which includes all 4 deg$^2$ of F2 and approximately 2.63 deg$^2$ of F5), we select stars by using the DLS catalogue parameter \textit{dlsqcprob}, which is a $\chi ^2$ comparison to the measured PSF on a given DLS stack, converted into a likelihood. Using the published system response from the NOAO website\footnote{http://www.noao.edu/kpno/mosaic/filters/} and the \citet{pickles} and \citet{bpgs} libraries of stellar spectra, we predict magnitudes for red stars which are significantly brighter than our actual measured magnitudes. The effect is more pronounced for redder stars, and while this colour dependence could be modelled using a classic linear colour term for stars, such a solution could not be expected to extend to galaxies, brown dwarfs, or QSOs, which might share the same $(r-i)$ colour but have quite dissimilar detailed spectra. It is therefore necessary to adjust the system response function in order to obtain correct colours for all objects simultaneously.

\subsection{Modelling the CCD response}

Of the four filters used in the DLS, only the $z$ filter shows signs of differing from its tabulated response curve. The $z$ filters used with MOSAIC and MOSAIC-II were long-pass filters whose transmission reaches 50 per cent of maximum near 8340$\textrm{\AA}$. For incident light of wavelength near 1$\mu$m, the system response is determined almost entirely by the response of the detector CCDs; the filter is essentially transparent at those wavelengths. The CCDs used in both MOSAIC cameras were thinned, back-illuminated silicon CCDs manufactured by SITe \citep{SiteCCDs}, the same detectors used in the Hubble Space Telescope's Advanced Camera for Surveys \citep{Clampin1998}. The absorption of silicon as a function of wavelength and temperature has been modelled for a number of different applications, including solar panels.

Absorption of visible light by silicon occurs by both direct and indirect processes. The direct process excites an electron into the conduction band in a single-photon interaction; the indirect process allows a photon from a wider range of energies to excite the electron in combination with a phonon. The necessity of the phonon in the silicon crystal makes the efficiency of the indirect process strongly temperature-dependent. The phonon can either contribute energy to the interaction or carry it away. We use the parametrization from \citet{Rajkanan1979}, who fitted model coefficients, $C_i$, to reproduce the direct and indirect band-gap absorption coefficients of silicon, $\alpha_{direct}$ and $\alpha_{indirect}$, as a function of temperature, $T$, in the form
\begin{equation}
\begin{split}
\alpha(T)=&\alpha_{direct}+\alpha_{indirect}\\
=&A_{direct}(\hbar\omega-E_{direct}(T))^{2}\\
 &+\sum_{i,j}C_{i}A_{j}(T)\left[\frac{\left\lbrace \hbar\omega-E_{gj}(T)+E_{pi}\right\rbrace^{2}}{e^{E_{pi}/k_{B}T}-1}\right.
\\  &+\left.\rho_{i}\frac{\left\lbrace\hbar\omega-E_{gj}(T)-E_{pi}\right\rbrace^{2}}{1-e^{-E_{pi}/k_{B}T}}\right] . 
\label{eqn:siliconabs}
\end{split}
\end{equation}
The subscript $i$ runs over possible phonon energies $E_{pi}$, while the $j$ subscript corresponds to different indirect band gaps $E_{gj}$. The first term inside the large square brackets corresponds to absorption of a phonon of energy $E_{pi}$, which process is allowed for any incoming photon whose energy lies in the range $\left\lbrace E_{gj}(T)-E_{pi}\right\rbrace\leq\hbar\omega\leq\left\lbrace E_{gj}(T)+E_{pi}\right\rbrace$, while the second term represents the emission of a phonon to carry away excess energy; this is only allowed for $\hbar\omega\geq\left\lbrace E_{gj}+E_{pi}\right\rbrace$. In Equation~\ref{eqn:siliconabs}, the terms must be allowed or disallowed manually, as they do not naturally vanish for forbidden energies. As a result, the function has discontinuities in all derivatives. The temperature dependence of the band gap energies was taken from \citet{Varshni1967}, and assumes the form
\begin{equation}
\label{eqn:gapenergy}
E_{g}(T)=E_{g}(0)-\left[ \frac{\beta T^{2}}{T+\gamma}\right] .
\end{equation}
\citet{Rajkanan1979} then determined the $C_i$ in Equation~\ref{eqn:siliconabs} and the optimal band gap energies by a least-squares fit to the NASA absorption data. In order to convert this $\alpha$ into a quantum efficiency, it is merely necessary to consider the fraction of photons absorbed in a CCD of thickness $d$:

\begin{align}
I&=I_{0}e^{-\alpha d}&&\text{(definition of\ } \alpha \text{)}\\
QE(T,\lambda)&=1-\frac{I}{I_{0}}=1-e^{-\alpha(T,\lambda)d}&&&
\end{align}

\begin{figure}
\centering
\includegraphics[width=.99\hsize]{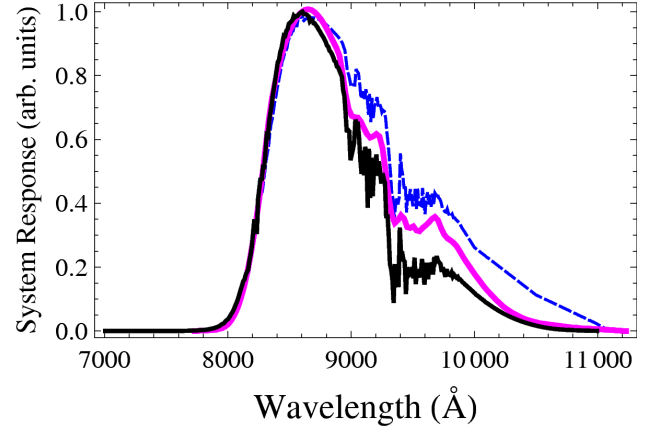}
\caption{System response functions used in our analysis: published MOSAIC $z$ filter and CCD plus KPNO atmosphere (blue dashed line), published SDSS $z$ response (magenta points), and our modified MOSAIC $z$ response (solid black line). MOSAIC response functions have been averaged over 10$\textrm{\AA}$ bins for plotting clarity. The modified MOSAIC response at 1$\mu$m has been reduced by 65 per cent from the published value. \label{filterfig}}
\end{figure}

We then calculate the quantum efficiency of a silicon detector of the design thickness of the SITe detectors (14--16$\mu$m) at the operating temperature of the MOSAIC cameras, which is determined from the FITS image headers to be 168K. This is combined with the filter transmission and the atmospheric transmission at KPNO to obtain a modified response function, shown in Figure~\ref{filterfig}. Atmospheric absorption is based on the atmospheric throughput of \citet[][]{atmo}\footnote{ftp://ftp.noao.edu/catalogs/atmospheric\_transmission}, and the overall depth of the absorption features has been allowed to vary as a free parameter in order to optimize the stellar colour fit. The modified MOSAIC $z$ response is slightly less red-sensitive than the SDSS $z$ response; this distinguishes it from the published MOSAIC $z$ response, which claimed a greater red sensitivity than SDSS. This difference in relative sensitivity leads to the qualitative difference in the colour trends of stars in Figure~\ref{starfig}. The contours in Figure~\ref{starfig} are a kernel density estimate made by summing a Gaussian error distribution (assumed uncorrelated) for each star. Although there are small discrepancies in zero-point ($\simeq 0.02$ mag) between DLS fields with respect to the predicted stellar locus, these are not relevant to the effect of the $z$-band response, which is colour-dependent. To remove the effect of the zero-point differences, the predicted SDSS~$z$~--~DLS~$z$ colours have been offset by $-0.016$ in F2 and $+0.024$ in F5 to ensure that zero colour for the library stars matches the mean of the measured star colours for $0<(r-i)<0.4$. The errors on the binned stellar colours are derived by first resampling each star from its own error distribution, then bootstrap resampling those stars to determine the standard deviation of the mean; the sampling was carried out 10,000 times on each of 10,000 error distribution resamples. This system response has now been adopted for all of the DLS stellar colour calibrations, as well as the photometric redshift estimation.

This correction to the MOSAIC $z$-band system response is essential to stellar science, stellar locus calibration, and photo-z's that use this filter. Using the published filter curves leads to disagreement between the blue and red stars of order 0.1 magnitudes in the zero-point of the $z$ filter, which leads to calibration instabilities several times larger than our typical systematic zero-point errors ($\sim$0.02 mag). For cool star and brown dwarf science, the differences are even more stark: for library L dwarf spectra \citep[the sample from Table 2 of ][]{ryan11}, the difference between the predicted colours using the published and modified sensitivities is as large as 0.35 magnitudes. The effect of a 0.1 magnitude error on photo-z's is likewise dramatic: as we will show in  \S\ \ref{tweak}, a 0.1 magnitude miscalibration of the $z$-band zero-point could lead to a systematic shift of 0.2 in redshift at fixed type; and worse, in practice the types would not be held fixed, leading to substantial misclassification and errors in the template tweaking process. We recommend the adoption of this response curve by anyone interested in using DLS $z$-band data, or by any other group using archival $z$-band data obtained with the original MOSAIC cameras.

\section{Photometric Redshifts}
\label{photoz}
With a corrected $z$ response function and proper photometry in hand, we now turn to computing the photo-z's.  With only four optical bands to constrain SEDs, we do not expect precise redshift estimates (the survey design goal was $\sim10$ per cent uncertainty in photo-z at median redshift).  We use version 1.99.3 of the publicly available\footnote{http://www.its.caltech.edu/$\sim$coe/BPZ/} code BPZ \citep[]{BPZ:00}, a template-based code with Bayesian priors on the apparent magnitude, to perform the redshift estimates.  We make adjustments to two components of the procedure in order to optimize photometric redshift performance: the template set used to calculate the initial likelihoods, and the Bayesian apparent magnitude and type prior applied for the final posterior analysis.  
  
One challenge in this procedure (which is also a factor for both the template modification and the prior determination) is that we must perform an initial classification of each galaxy in the training set to a particular galaxy type.  We do this by finding the ``best-fitted'' template to the broad band fluxes at the fixed spectroscopic redshift when compared to those expected from an initial discrete set of SEDs.  This is problematic for two reasons: first, because galaxies do not easily separate into discrete types, showing instead a more continuous distribution of properties; and second, because evolution of the galaxy population as a function of redshift may further blur the partitioning of our training sample, and hence our resulting template and prior results.  In nearly all current photometric redshift implementations the templates are assumed not to evolve, and model fluxes are determined by simply redshifting the static SED.  In this paper we will assume a discretized template set and no evolution in the templates, but we discuss possible modifications in Section~\ref{conclusions}.
For both the template adjustment and prior training, we find the best-fitting type using the built-in $\textrm{ONLY\_TYPE}$ setting of the BPZ package. This option constrains the redshift to the correct spectroscopic value and finds the best-fitting template given the four observed DLS fluxes. Note that because we fix the redshift to the correct value, the best-fitting type may be different from the type assigned in the full photo-z fit, which allows the redshift to vary across the full redshift range when determining the type.

\subsection{Tweaking the SED Templates}\label{tweak}

Following \citet[]{BPZ:00}, the heart of template-based photometric redshift codes is a simple $\chi^{2}$ calculation comparing the observed fluxes of a galaxy with model fluxes found by convolving a model SED with the system throughput with $N$ passbands:

\begin{equation}\label{chisqeqn}
\chi^{2} = \sum_{i=1}^{N}\frac{(f_{O,i}-\alpha\,f_{T,i}(z))^{2}}{\sigma_{i}^2}
\end{equation}

\noindent where $f_{O,i}$ is the observed flux in the $i$th filter, $f_{T,i}(z)$ is the convolved flux of the SED model template at redshift $z$, $\sigma_{i}$ is the flux error, and $\alpha$ is a scaling factor.  Note that the use of the $\alpha$ factor to scale all fluxes implicitly assumes neither luminosity nor redshift dependence of galaxy SEDs, only a simple scaling -- an assumption that is very likely wrong in practice.  We will discuss this further in section~\ref{conclusions}.  If we assume that our template set roughly spans all possible galaxy types, and we test on a grid of redshifts that covers all possible redshifts, then we can assign a likelihood ${\mathscr L}\,\propto\,{\textrm exp}(-\chi^{2}/2)$ and normalize the total likelihood to one.  We then marginalize over galaxy type (by simply summing the individual type likelihoods) to calculate the one-dimensional redshift probability density function (PDF), p(z).  The choice of model SED set is of paramount importance in this calculation.  The choice of SED set directly determines the PDF returned by the algorithm.  Any differences between the model SEDs chosen to represent the galaxy population and the true population will manifest as a bias in the predicted redshift distribution.  We can adjust our template SEDs to better match a set of training data, a process that is often referred to as ``tweaking'' the templates.

The template correction procedure is straightforward, and similar to the procedure described briefly in \citet[]{Ilb:06}.  We take advantage of the fact that our fixed filter curves sample different wavelength ranges of an SED at different redshifts.  In fact, given a set of galaxies with known spectroscopic redshift and classified by SED, we can effectively fix the SED at redshift $z=0$ and blueshift the filter curves to check which portion of the SED is being sampled.  We calculate the effective wavelength, $\lambda_{eff}$, of each filter as the weighted mean of the filter transmission curve $T(\lambda)$:  

\begin{equation}\label{leffeqn}
\lambda_{eff}=\int{\lambda\,T(\lambda)\,d\lambda}
\end{equation}

We perform an initial classification of each galaxy using its known spectroscopic redshift from a training set (here, the SHELS redshifts) to the best-fitting template from the six default templates for older versions of BPZ, consisting of the four CWW SEDs, plus the \citet[]{Kin:96} SB3 and SB2 ``starburst'' templates (collectively we will refer to this set as the CWW+SB template set), and assign the scaled flux in each filter to the wavelength 
\begin{equation}\label{shiftleffeqn}
\lambda_{eff,z}=\frac{\lambda_{eff}}{1+z_{s}} 
\end{equation}
\noindent Individually, the measurements are quite noisy. We take the mean flux in bins of width $\approx75-100\textrm{\AA}$ to obtain a moderate-resolution adjusted SED for each initial template type.  If fewer than ten flux points are present in the bin we linearly interpolate between adjacent bins.

Although the procedure could be performed iteratively, reclassifying the training set with the modified SED set, in practice the classifications converge very quickly, and the resultant SEDs are nearly unchanged on subsequent iterations.  We do not tweak the two starburst templates, both because we have very few blue galaxies in the magnitude limited SHELS sample, and because our wide bins and broad filters would smooth out emission lines present in the spectra.  The SHELS data only cover $0\,\leq\,\textrm{z}\,\lesssim\,0.8$ and given the effective wavelengths of our filter set we can only tweak the wavelength range $\sim2500-8500\textrm{\AA}$.  Beyond this range we simply use the original template SED, appropriately scaled to match the tweaked portion of the spectrum.

\begin{figure}
\centering
\includegraphics[width=.99\hsize]{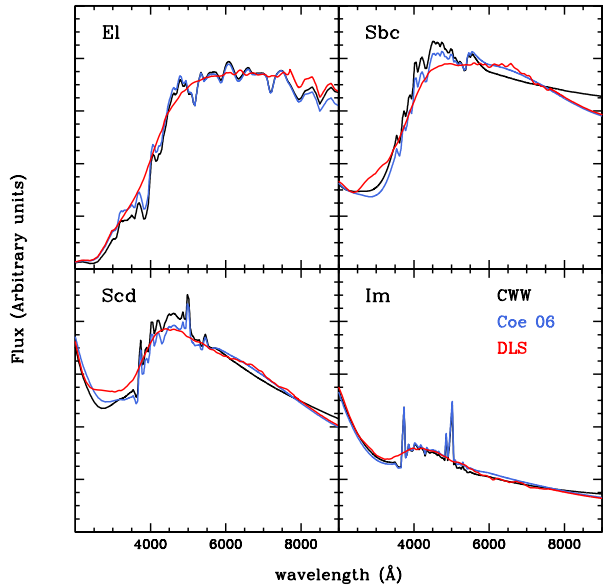}
\caption{Comparison of the original CWW templates (black) to the DLS tweaked templates (red).  The templates of \citet[]{colorpro} are shown in blue for comparison.  While the modified templates smooth over some small-scale spectral features, the resulting colours are a better match to those observed by DLS for the SHELS data set.
  \label{tweaksed}}
\end{figure}

Figure~\ref{tweaksed} shows the modified SEDs compared to both the original CWW+SB templates and the templates from \citet[]{colorpro}.  One natural consequence of taking median values of the fluxes measured in our broad-band filters, which effectively act as broad smoothing kernels, is the loss of small scale spectral features of the original templates.  While this results in some loss of rigor in our templates, they will have better predicted colours compared to the CWW+SB templates, resulting in less bias and scatter in our photometric redshift predictions.  Figure~\ref{tweaksed} also shows that our templates agree well qualitatively with those of \citet[]{colorpro}, particularly at the red end.

\subsection{Prior Determination}\label{prior}

The BPZ software package used to calculate photometric redshifts for DLS includes the application of an apparent-magnitude-based Bayesian prior on both galaxy type and redshift.

The default prior is based on a small set of data from the Hubble Deep Field North \citep[HDFN;][]{Will:96}.  The data set used to train the prior consists of 737 spectroscopic redshifts over the relatively small area of the HDFN, where sample variance may be a problem.  It also treats all galaxies brighter than $m<20$ as if they were 20th magnitude, which can lead to bias for bright objects.  In order to examine differences from the HDFN prior and obtain a more accurate prior at brighter magnitudes, we fit a new prior to the same parametrization used by BPZ, using both the DLS SHELS data and the public VIMOS VLT Deep Survey (VVDS) deep spectroscopic survey \citep[]{vvds:05}.  We use the DLS data set to constrain the prior for galaxies brighter than $R<21.25$ (Vega) and the VVDS Deep for fainter galaxies.  The SHELS galaxies are treated in the same manner described in Section~\ref{tweak}.  The VVDS Deep first epoch release catalogue contains 11564 spectra in a 0.61~deg$^{2}$ area.  While still limited in terms of areal coverage and subject to sample variance, this data set provides a larger area with more spectroscopic redshifts than the HDFN.  After cuts for quality flags and spatial matching we are left with 8145 galaxies, which are assigned a best-fitting template by fixing the correct spectroscopic redshift and finding the best-fitting UBVRI \citep[]{McCvvds:03} colours.

 We use the DLS $R$-band as the reference magnitude and, following the convention of the DLS, we work in Vega magnitudes, as this is the system that best matches the $BVR$ filter set.  In order to place the VVDS $r$-band magnitudes on the same zero-point scale as the DLS, we cross match VVDS with the $r$-band magnitudes from CFHTLS Field D1, as CFHTLS $r$-band has the same zero-point as the SDSS $r$-band\footnote{See http://terapix.iap.fr/rubrique.php?id\_rubrique=252}.  A simple cross-match of the surveys gives $r_{CFHT} \approx r_{VVDS} + 0.075$ and $r_{SDSS} \approx R_{DLS} + 0.14$ (which includes the $0.20$ AB-to-Vega conversion factor for $r$-band), for an approximate relation $r_{VVDS} \approx R_{DLS} + 0.065$, which we incorporate into the value used for the VVDS $m_{0}$ value in the prior.

\citet[]{BPZ:00} assumes independence of the type and redshift portions of the prior:
\begin{equation}
p(z,T|m_{0}) = p(T|m_{0})p(z|T,m_{0}) 
\end{equation}
\noindent where T is the galaxy type, divided into three broad types as in BPZ as Elliptical (the ``El'' template), Spiral (the ``Sbc'' and ``Scd'' templates), and Irregular (the ``Im'', ``SB3'', and ``SB2'' templates), and $m_{0}$ is the apparent magnitude of the galaxy in a reference band.  The galaxy type fraction is parametrized as an exponential in type fraction as a function of apparent magnitude:
\begin{equation}
p(T|m)=f_{T}e^{-k_{t}(m-m_{0})}
\end{equation}
\noindent where $f_{T}$ is the fraction of galaxies of type T at the reference magnitude, $m_{0}$. 
The type redshift prior is parametrized as:
\begin{equation}
p(z|T,m) \propto z^{\alpha_{T}}exp\left\{-\left( \frac{z}{z_{0T}+k_{mT}(m-m_{0})}\right)^{\alpha_{T}}\right\}
\end{equation}
\noindent where $\alpha_{T}$ is related to how peaked the redshift distribution of type T is, $z_{0T}$ and $k_{mT}$ control how quickly the peak redshift changes as a function of magnitude, and the proportionality is used to scale the integral of the summed probabilities to unity for proper normalization.
To ensure proper normalization, $k_{T}$ is undefined for the Im/SB type, and is assumed to equal the fraction unassigned to the El and Sp types.  For further detail on the prior form, see \citet[]{BPZ:00}.

\begin{figure}
\centering
\includegraphics[width=.99\hsize]{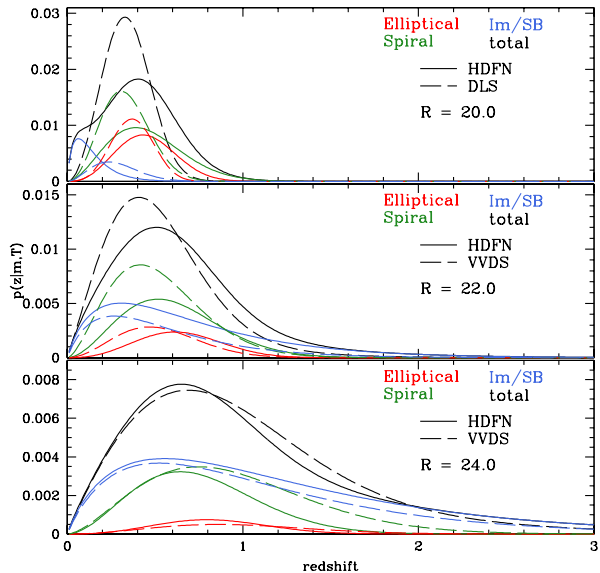}
\caption{Comparison of the HDFN prior (solid lines) to the DLS/VVDS prior (dashed lines) for Elliptical (red), Spiral (green), Im/SB (blue) and total (black).  The priors are markedly different at bright magnitudes, but are similar at $R=24$.  The differences at bright magnitudes may be due to sample variance or the extra freedom afforded by fitting to two data sets.
  \label{priorfig}}
\end{figure}

By parameterizing the prior with two data sets (SHELS for $R<21.25$ and VVDS for $R \geq 21.25$) we do introduce a possible discontinuity in the prior prediction at $R=21.25$.  We have taken care to minimize this effect by constraining the initial fractions, $f_{T}$, of VVDS to match the parametrized fraction predicted from SHELS.   Table~\ref{prior_table} lists the best-fitting parameters for both the SHELS and VVDS priors.  The fit is performed with a basic downhill simplex algorithm, and errors are estimated at the minimum assuming a multidimensional Gaussian in likelihood.  The parameters $k_{T}$ and $f_{T}$ are fit simultaneously, as are $\alpha_{T}$, $z_{0T}$ and $k_{mT}$.  While the best-fitting parameters are a good description of the data, we caution that the likelihood maximization occurs in very large five- and nine-dimensional likelihood spaces, so the errors are almost certainly underestimated.  Figure~\ref{priorfig} shows a comparison of the HDFN prior to the DLS/VVDS prior at three representative magnitudes.  The priors are markedly different at bright apparent magnitudes; for instance, the VVDS sample predicts a larger fraction of spiral galaxies compared to HDFN.  This may be due to the extra freedom allowed in our prior when we fit $R < 21.25$ with DLS SHELS data and $R > 21.25$ with VVDS data, giving two times the number of free parameters as were used in the HDFN prior.  It may also be due to sample variance or Poisson fluctuations in the number of bright galaxies in the limited area of the two spectroscopic surveys in question, particularly the very small HDFN.
At $R=24$ the two priors are quite similar; however, the VVDS prior has a slightly larger tail at high redshift than HDFN, and this difference grows at fainter magnitudes.  Once again, this could be due to sample variance or to the extra freedom in the DLS/VVDS parametrization.  We will discuss further investigation of the prior in Section~\ref{conclusions}.

\begin{table*}
\begin{center}
\caption{Best-Fitting Prior Parameters}
\begin{tabular}{|l|c|c|c|c|c|}
\hline
\hline
Type & $f_{T}$ & $k_{T}$ & $\alpha_{T}$ & $z_{0T}$ &$k_{mT}$\\
\hline
 &  & DLS & $m_{0}=18.0$ & &\\
\hline
El & 0.55$\pm$0.02 & 0.25$\pm$0.010 & 2.99$\pm$0.091 & 0.191$\pm$0.0050 & 0.089$\pm$0.0033\\
Sp & 0.39$\pm$0.02 & -0.175$\pm$0.011& 2.15$\pm$0.058 & 0.121$\pm$0.0055 & 0.093$\pm$0.0040\\
Im/SB & 0.06$\pm$0.02 &  & 1.77$\pm$0.11 & 0.045$\pm$0.0065 & 0.096$\pm$0.014\\

\hline
 &  & VVDS & $m_{0}=21.25$ & & \\
\hline
El & 0.25$\pm$0.02 & 0.565$\pm$0.028 & 1.957$\pm$0.165 & 0.321$\pm$0.028 & 0.196$\pm$0.016\\
Sp & 0.61$\pm$0.02 & 0.155$\pm$0.013 & 1.598$\pm$0.08 & 0.291$\pm$0.016 & 0.167$\pm$0.010\\
Im/SB & 0.14$\pm$0.02 &  & 0.964$\pm$0.045 & 0.170$\pm$0.012 & 0.129$\pm$0.013\\
\hline
\label{prior_table}
\end{tabular}
\end{center} 
\end{table*}

\section{Photometric Redshift Results}\label{results}

In order to test our photometric redshift results with the new templates and prior, we require a completely independent test sample with known spectroscopic redshifts.  This is normally done by setting aside randomly selected galaxies from the training sample that are not used in the training procedure and calculating their redshift probabilities.  While this procedure may be valid, it does result in a test sample that is subject to the same selection criteria, photometric offsets, and other possible systematic effects as the training sample.  A completely independent test set is more desirable, provided that it is complete to sufficient magnitude limits.

The PRIMUS spectrocopic data set described in Section~\ref{spectro} provides exactly this, with 9107 spectroscopic redshifts, 703 of which are at $R>23.3$, measured in a field that is spatially distinct from those used in the training of the templates and prior.  As a further benefit, this sample resides in DLS Field F5, observed with the Blanco telescope at CTIO, whereas the SHELS sample used to train both the SEDs and prior is located in Field F2, observed with The Mayall telescope at KPNO.  The distinct nature of these two samples enables us to test the photo-z performance in the presence of any field-to-field systematics that might be present in the survey.

\begin{figure*}
\centering
\includegraphics[width=.99\hsize]{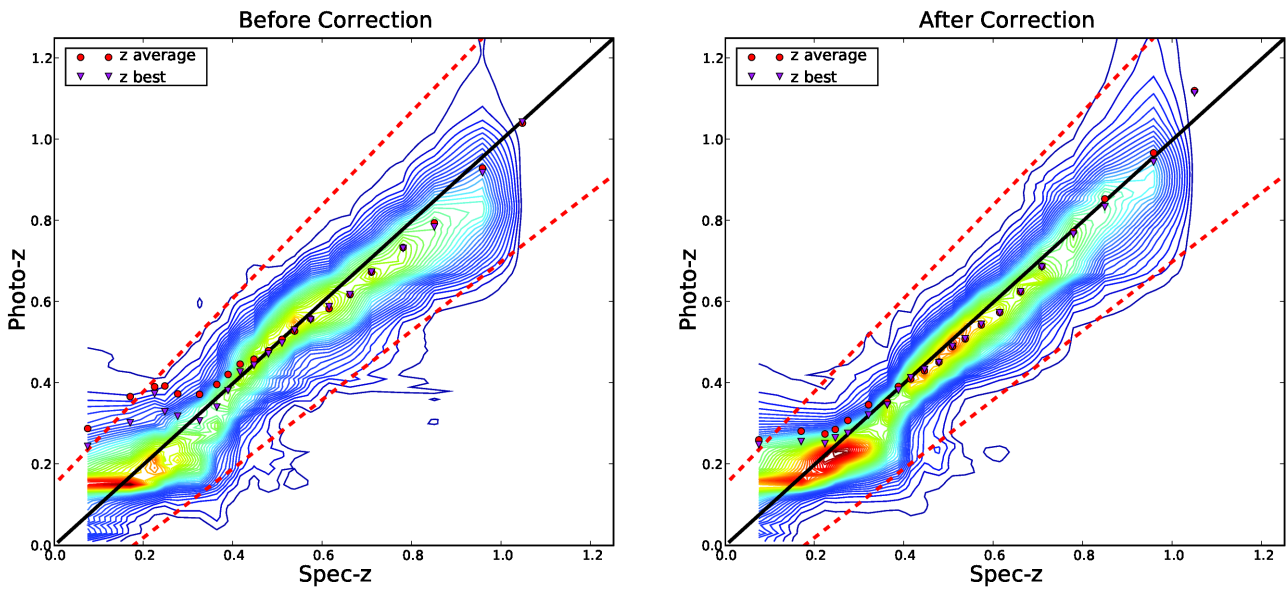}
\caption{Plot of the summed probability distribution functions of the DLS PRIMUS galaxies as a function of spectroscopic redshift using the original CWW+SB filters and HDFN prior (left) and the tweaked filters and DLS/VVDS prior(right).  Contours are linearly spaced and reflect both the number of galaxies and the probability in each redshift interval.  Also shown are the mean single point photo-z estimates in twenty bins for both the peak likelihood estimate (z best, purple triangles) and average probability per galaxy (z mean, red circles).  Photo-z bias and scatter are reduced by the template and prior optimizations.  The threshold that defines the catastrophic outlier fraction are shown as red dashed lines defined by $0.15(1+z)$.  The optimization reduces overall redshift scatter by $\approx$20 per cent at $z < 0.6$ and reduces bias by a factor of two at $z>0.6$.  In the improved results the point estimators show a pronounced bias at $z<0.25$ though this bias can be mitigated by using the full photo-z PDF.
  \label{szpz_plot}}
\end{figure*}

Figure~\ref{szpz_plot} shows two contour plots of the probability density p(z) as a function of spectroscopic redshift for the 9107 PRIMUS galaxies in our test sample.  Fifty linearly spaced contours are shown, encoding both the density of spectroscopic objects at each redshift and the spread in the PDFs.  Red dashed lines indicate the threshold that defines ``catastrophic outliers" where the photo-z differs from the spec-z by $\pm 0.15(1+z)$. The left panel uses the unmodified CWW+SB templates and default HDFN prior, while the right panel uses the tweaked templates discussed in Section~\ref{tweak} and the DLS/VVDS prior discussed in Section~\ref{prior}. 

Also shown for comparison are binned averages for two ``point'' estimates of photo-z: triangles represent ``z best'' (Z\_B in BPZ), the peak likelihood value for each galaxy, while circles represent ``z average,'' the weighted mean value of p(z) for each galaxy.  We see excellent agreement between the spectroscopic redshifts and all three photometric redshift estimators for $z>0.3$.  Our shortest-wavelength band, $B$, with its effective wavelength of 4420$\textrm{\AA}$, does a poor job of constraining the 4000$\textrm{\AA}$ break (the dominant feature for photo-z determination) at low redshift.  This deficiency manifests itself as an increase in photo-z scatter at $z<0.25$ for the p(z) distribution.  The single-point estimators show a pronounced bias at low redshift, as a single number is unable to capture the the asymmetric PDFs for these galaxies.  We strongly discourage use of the single point ``z best'' estimators \citep[see, e.~g.,][]{Abr:11}.

 The quality of photometric redshifts is often summarized by three parameters: the (assumed Gaussian) ``width'' of the distribution ($\sigma_{z}$), the redshift bias, and the catastrophic outlier fraction.  We define the redshift bias as the mean value of the difference between the photo-z and spec-z estimates.  The fraction of ``catastrophic outliers'' is defined as the fraction of galaxies with redshift estimates different from the spectroscopic redshift greater than some fixed amount. 
For the point estimate photo-z's, $\sigma_{z}$ uncertainty is calculated as:
\begin{equation}
\frac{\sigma_{z}}{1+z} =  \sqrt{\frac{1}{N}\sum_{i=1}^{N}{ \frac{(z_{p,i}-z_{s,i})^{2}}{(1+z_{s,i})^{2}}}}
\end{equation}
\noindent and the catastrophic outlier fraction is defined as the fraction of galaxies with  $|z_{p} - z_{s}|>0.15(1+z_{s})$.  For the full p(z), to measure $\sigma_{z}/(1+z)$ we sum $p(z) - z_{s}$ for all galaxies in a redshift interval and fit a Gaussian of width $\sigma$.  For the catastrophic outlier fraction, we simply sum the amount of probability outside of $|0.15(1+z_{s})|$.

\begin{figure}
\centering
\includegraphics[width=.99\hsize]{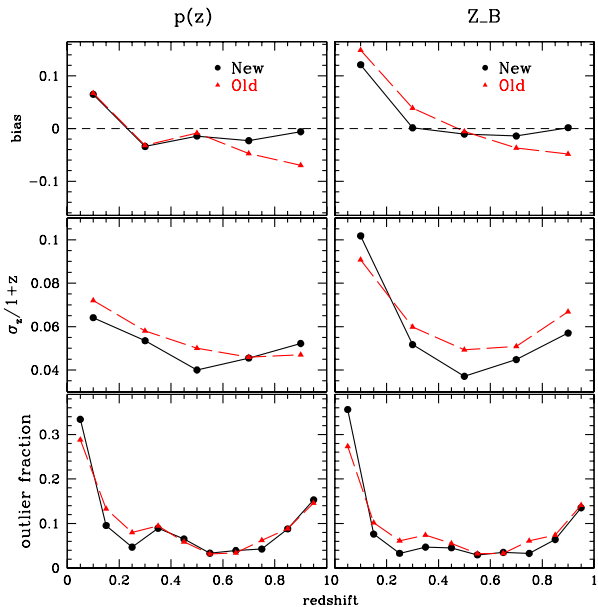}
\caption{Measured photo-z bias (top panels), photometric scatter ($\sigma_{z}/(1+\textrm{z})$, middle panels) and catastrophic outlier fraction (bottom panels) as a function of spectroscopic redshift for the PRIMUS galaxies in DLS computed for both the original CWW+SB templates and HDFN prior (red) and tweaked templates and DLS/VVDS prior (black).  The left panels show results using the full p(z) and the right panels show the Z\_B point photo-z estimator.  Improvements in the templates and prior reduce scatter by 20 per cent and dramatically reduce bias at higher redshift.  
  \label{sigma_catoutlier}}
\end{figure}

Figure~\ref{sigma_catoutlier} shows summary statistics in spectroscopic redshift bins for the PRIMUS dataset calculated for photo-z's generated with the original CWW+SB templates and HDFN prior and the updated templates and DLS/VVDS prior.  The left panels show estimates based on the redshift PDFs, right panels show estimates using the $Z\_B$ point estimate. 
Use of the updated prior and template set led to a $\sim20$ per cent reduction in photo-z scatter at $\textrm{z}<0.6$ and a reduction in photo-z bias by a factor of two at $\textrm{z}>0.6$.
All estimators show a large bias ($>0.06$) at $\textrm{z}<0.2$.  This is not unexpected, as we do not allow the photo-z estimators to assign redshifts below $\textrm{z}_{p}<0.01$, which naturally leads to a truncation bias at low redshift.  The lack of broad spectral features (e.g., the 4000$\textrm{\AA}$ break) and the presence of degeneracies in the DLS colours (which will be discussed in the following subsection) both contribute to this bias, as well as to larger uncertainties in the photo-z estimates.  Use of the full p(z) does result in better performance than the point estimate in the $0 \le \textrm{z} \le 0.2$ range, with an average uncertainty of 0.064 for p(z) versus 0.102 for the point estimates.  However, due to the large scatter we do not recommend the use of DLS photometric redshifts below Z\_B$\lesssim0.25$.  The catastrophic outlier rate remains nearly unchanged, as the effects of reduced bias and scatter occur largely within the very broad interval defined by our $>0.15(1+\textrm{z}_{s})$ catastrophic outlier criterion.

Measuring the resulting scatter in the photo-z versus spec-z relation yields an estimate for the actual uncertainty in measuring photometric redshifts (at least for those objects in the test sample).  However, this statistical measure does not tell us anything about the reliability of the redshift probability densities estimated for each galaxy.  In calculating the redshift PDF we assume, among other things, that our template set is complete and spanning (and, in nearly all current photo-z codes, that the templates do not evolve).   We also assume that our prior is correct, even when extrapolated to fainter magnitudes than are covered by the training set.  The breakdown of any of these assumptions can lead to an incorrect estimate of the redshift PDF.  In other words, we want to ask: do the  redshift uncertainties based on the posterior probability distributions for individual galaxies match the actual measured uncertainties?

To test the accuracy of our p(z) estimates for the PRIMUS data, we determine the redshift bounds containing a fixed fraction of each individual galaxy's total probability (extending from ``zmin'' to ``zmax'' and centred on the peak probability, $Z\_B$) and and compare this to the actual fraction of spectroscopic redshifts that fall within this interval.  This is very similar to a procedure described in  \citet[]{fern-soto02}.  Table~\ref{pz_table} lists the probability fraction and the number and fraction of galaxies that fall within the corresponding ``zmin'' to ``zmax'' interval for all PRIMUS galaxies between $0.01 \leq z \leq 1.0$.  If the p(z) estimates are accurate then we expect these fractions to agree.  For example, if our $1\sigma$ estimates determined from p(z) are correct, then 68 per cent of the spectroscopic redshifts should fall within these $1\sigma$ bounds.  The values are somewhat noisy, as the DLS photo-z algorithm samples the redshifts at intervals of $\Delta\,\textrm{z}=0.01$, so p(z) is often localized to a small number of non-zero values.  To mitigate this resolution effect, we linearly interpolate the PDF to a grid of $\Delta\,\textrm{z}=0.0025$.  
We test the addition of a redshift cut of $0.25 \leq \textrm{z} \leq 1.0$ to avoid the biased p(z) predictions at low redshift. We also select galaxies based on the BPZ ODDS parameter, which is a measure of the amount of probability within a fixed interval of $0.06(1+\textrm{z})$ around the Z\_B value.  BPZ ODDS selects galaxies with peaked and unimodal p(z) distributions.  The observed fractions when restricting to $0.25 \leq \textrm{z} \leq 1.0$ and BPZ ODDS $> 0.6$ differ by less than one per cent compared to the total sample shown in Table~\ref{pz_table}.
There is some evidence that we are underestimating the core of the distribution (as 60 per cent of galaxies actually lie within the 50 per cent confidence intervals), and are overconfident in our estimate of the distribution tails (e.g., the 99 per cent confidence region contains only 92 per cent of galaxies). However, Table~\ref{pz_table} shows that the p(z) fractions agree to within 10 per cent with the number of spec-z objects and thus the p(z) are approximately representative of the true redshift uncertainties for this data set.

\begin{table}
\begin{center}
\caption{Predicted and Observed Probability Density Fractions}
\begin{tabular}{|c|c|c|}
\hline
p(z) Fraction & N in interval & Observed Fraction\\
\hline
0.30 & 3509/9011 & 0.39\\
0.50 & 5419/9011 & 0.60\\
0.68 & 6846/9011 & 0.76\\
0.90 & 8014/9011 & 0.89\\
0.95 & 8199/9011 & 0.91\\
0.99 & 8320/9011 & 0.92\\
\label{pz_table}
\end{tabular}
\end{center} 
\end{table}

\subsection{Type Dependence}\label{types}
\begin{figure}
\centering
\includegraphics[width=.99\hsize]{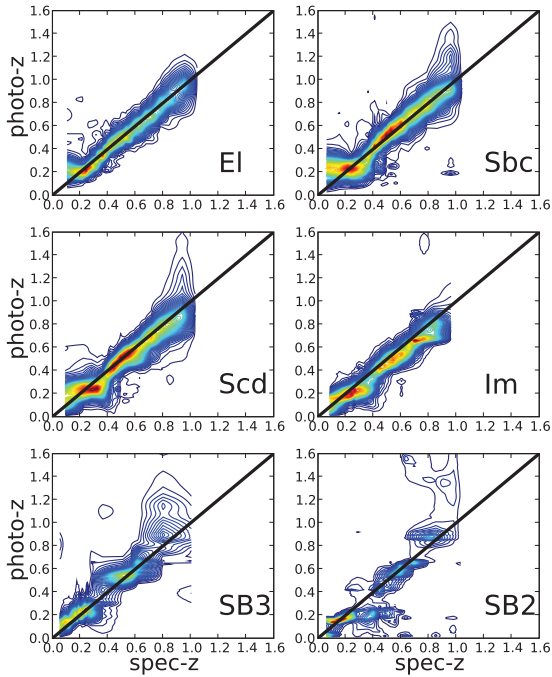}
\caption{The summed photometric redshift PDFs as a function of spectroscopic redshift for each of the six galaxy templates.  The Elliptical and Spiral types show a strong bias at $z < 0.25$ but excellent performance  at intermediate redshifts.
  \label{type_szpz}}
\end{figure}

We now examine the photo-z's of the PRIMUS sample, broken up by galaxy type.  We do not attempt to classify the galaxies from the spectra themselves; 
rather, we use the best-fitting type found with BPZ's $ONLY\_TYPE$ option briefly described in Section~\ref{photoz}. 
Figure~\ref{type_szpz} shows the summed PDFs as a function of PRIMUS spectroscopic sample from $0 \leq z \leq 1$ broken into types. As we run BPZ with $\textrm{INTERP=2}$, each type also includes the adjacent interpolated templates.  The starburst templates, SB3 and SB2, appear particularly noisy, partially due to a smaller number of galaxies (only 472 and 668 galaxies classified as SB3 and SB2 respectively) although their less extreme colours, due to a smaller 4000$\textrm{\AA}$ break, also lead to increased scatter.

\begin{figure}
\centering
\includegraphics[width=.99\hsize]{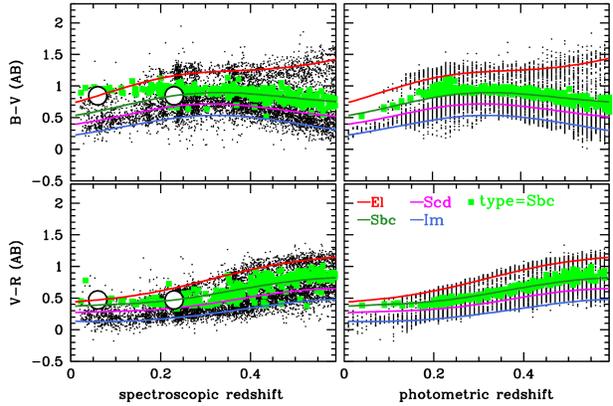}
\caption{Observed $B-V$ (top) and $V-R$ (bottom) colours as a function of spectroscopic redshift (left) and photometric redshift (right) for the PRIMUS galaxies.  Model template colour tracks are shown for the El (red), Sbc (green), Scd (magenta) and Im (blue) templates.  The green points show galaxies classified as type Sbc, which should match the green Sbc colour track.  Departures from the expected colour tracks are evident in the spectroscopic redshift plot, and almost no E, Sbc, and Scd galaxies are assigned to $z<0.2$ due to colour degeneracies.  A specific degeneracy between El at $z=0.06$ and Sbc at $z=0.23$ is highlighted with black and white dots.
  \label{szpz_color}}
\end{figure}

\begin{figure}
\centering
\includegraphics[width=.99\hsize]{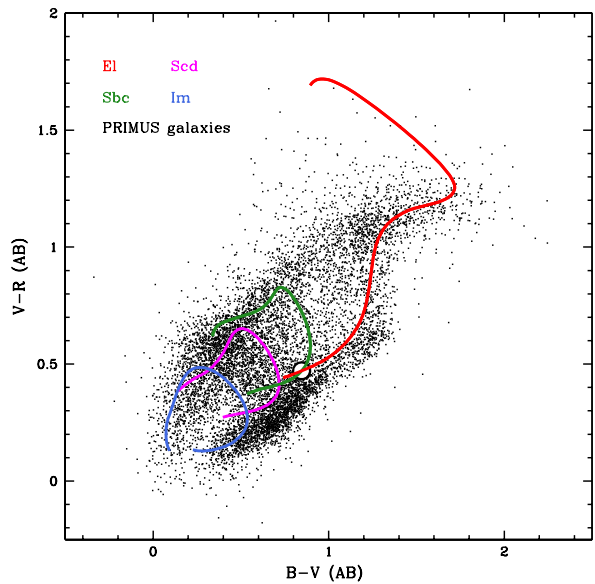}
\caption{Colour-colour plot for the PRIMUS galaxies in DLS.  Model template colour tracks are shown for the El (red), Sbc (green), Scd (magenta) and Im (blue) templates for $z=0.0$ to $z=1.1$.  The areas of overlap between the model tracks show that significant degeneracies exist at low redshift for the DLS filter set.  The particular degeneracy highlighted by the grey stars in Figure~\ref{szpz_color} is indicated here by the black and white dot at ($B-V$)$=0.85$ and ($V-R$)$=0.47$.
  \label{plot_colcol}}
\end{figure}

The pronounced bias and larger uncertainties at z$<0.2$ are noticeable for the Elliptical and Spiral galaxies.  These are due to colour degeneracies in the DLS filter set.   Figure~\ref{szpz_color} shows the observed ($B-V$) (top panels) and ($V-R$) (bottom panels) colours as a function of spectroscopic redshift (left panels) and photometric redshift (right panels) for the PRIMUS galaxies.  
The red, green, magenta, and blue lines show the E, Sbc, Scd, and Im model template colours respectively.
Green points highlight the galaxies classified as Sbc (left).  As expected, the Sbc galaxies match the predicted colours to the best-fitting type at the assigned point photometric redshift very well.  However, several prominent areas of departure of the Sbc galaxies from the proper colour track are present, particularly the ($B-V$) colour at $\textrm{z}<0.2$, and almost no galaxies are assigned to types E, Sbc, or Scd at $\textrm{z}<0.2$.  For the DLS filter set, there is a degeneracy between expected colours of the templates at $\textrm{z}\sim0.0-0.1$ and the next-bluer template at $\textrm{z}\sim0.2-0.3$.  
A particular degeneracy is highlighted by the black and white dots, which show that the Elliptical at $\textrm{z}=0.06$ and Sbc at $\textrm{z}=0.23$ have nearly identical colours. ($R-z$) is omitted for brevity, but does not offer much discriminating power, with template values of ($R-z$)$=0.59$ for the El and ($R-z$)$=0.57$ for the Sbc at the redshifts in question).
This is even more evident in Figure~\ref{plot_colcol}, which shows the ($V-R$) versus ($B-V$) colour-colour plot for the PRIMUS galaxies in DLS.  Model colours for El (red), Sbc (green), Scd (magenta) and Im (blue) are shown from $\textrm{z}=0.0$ to $\textrm{z}=1.1$.  There is significant overlap in the model colour tracks for the E, Sbc, and Scd galaxies, particularly at low redshift.  The degeneracy from Figure~\ref{szpz_color} is indicated by the black and white dot.
Such overlaps in colour-colour space show that the mapping from colour to redshift is not unique, leading to degenerate solutions and broad redshift PDFs.  The addition of extra colours (for instance, adding $u$-band or near-infrared data to the DLS) could break such degeneracies and improve low redshift photo-z results.

\subsection{The DLS Galaxy Redshift Distribution}\label{dls_sample}

\begin{figure}
\centering
\includegraphics[width=.99\hsize]{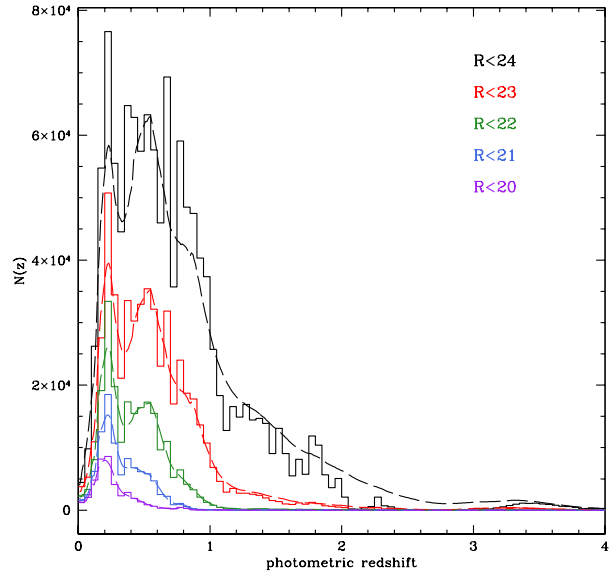}
\caption{Photometric redshift N(z) summed over all five DLS fields as a function of R magnitude.  Solid histogram shows the single point Z\_B estimator, while the dashed curves show the sum of the p(z) estimates.  The summed p(z) distributions provide a smooth representation of N(z) and correctly account for the long-tailed and bimodal shape of the posterior distributions, particularly at $z > 2$.
  \label{bignz}}
\end{figure}

Rather than restricting ourselves to a spectroscopic subsample, we now show samples for the full DLS data set, similar to what would be used in a typical science case, though the exact selection will depend on the specific needs of the analysis.  These photometric redshifts have been used in several DLS-based science results, and each specific data set was subject to tests for systematic errors; see the individual papers for details \citep[]{Mor:12,Jee:12,Choi:12}.
Figure~\ref{bignz} shows the N(z) for apparent-magnitude-limited slices consisting of all five DLS fields.  The sample has been trimmed based on bright object and bad pixel masks, and a star/galaxy cut has been made based on the DLS {\it dlsqcprob} parameter.  No cuts have been made on the BPZ ODDS parameter, which measures the amount of posterior probability within a fixed distance of the peak. The solid histogram represents the single point Z\_B estimator, while the dashed curve shows the sum of the photo-z PDFs.  Comparison of the point estimate to the summed p(z) estimate shows qualitative agreement between the point and p(z) estimators for bright magnitudes, but increasing differences as fainter apparent magnitudes are included, particularly when the sample extends to $\textrm{z} \geq 2$.  The excess probability at $\textrm{z} \ge 2$ is due to both broadly peaked PDFs for galaxies at $\textrm{z}\sim2.0$ and contributions of secondary ``catastrophic outlier'' peaks in PDFs at $\textrm{z}\sim0.2-0.4$ where the colour degeneracy between the Lyman break and 4000$\textrm{\AA}$ break produces bimodal probability distributions.  As samples push fainter in apparent magnitude and include more broad-tailed and multi-modal PDFs, use of the full p(z) distribution is essential to obtaining accurate redshift estimates.  As is discussed in \citet[]{Jee:12}, using the sum of p(z) values also results in a much smoother form for N(z) that is well-defined on the grid, obviating the need to fit a parametrized form for N(z) that may be necessary if using a point estimate for some science applications.

Figure~\ref{bignz} also shows a distinct bimodal shape for the N(z).  The excess probability at $\textrm{z}\sim 0.25$ is due to the degeneracies shown in Figure~\ref{szpz_color}, where galaxies at lower redshift are degenerate with $\textrm{z}\sim0.25$ galaxies with slightly bluer colours, similar to the degeneracy discussed in Figure~\ref{szpz_color}.  This feature is specific to the $BVRz$ filter set and lack of constraint on the 4000${\textrm \AA}$ break at low redshift.  Even use of p(z) for these galaxies does not fully correct the problem.  We once again caution against using DLS photometric redshifts below Z\_B$\lesssim 0.25$.

\section{Conclusions And Future Work}\label{conclusions}

To optimize the quality of the photo-z's, we have improved several of the inputs to the photo-z calculation, including the response curves, the prior probabilities, and the spectral templates.  It is very important that proper photometry be obtained before adjusting the template SEDs, fitting for a new prior, or calculating photo-z's.  Any bias in the input photometry will mix systematic effects with the underlying relations that we are attempting to recover, often for specific subsets of the data.  For example, Figure~\ref{starfig} shows that using the incorrect $z$-band filter curve gives accurate expected colours for blue stars, but results in incorrectly predicted $z$-band magnitudes of more than 0.1 magnitudes for red stars.
Any such bias will propagate through the calculation in the form of both misestimated observed fluxes and incorrect galaxy type classifications and can adversely affect photo-z performance.  Thus, careful photometric calibration is of the utmost importance for photometric redshift surveys.

By replacing the published filter$+$CCD response for the MOSAIC camera's $z$ band with a new CCD response based on the absorption properties of silicon, we improved the predictions of red star colours by 0.1 magnitudes relative to SDSS. Importantly, this was accomplished with a complete revision of the response curve, to be used on all objects in the DLS, rather than a linear colour correction, which would have been calculated using only stars, and may not have fully corrected galaxy photometry.  The filter curve is also used in computing the expected model fluxes in the photo-z algorithm, so that by correcting  the filter curve itself we have eliminated a troublesome source of bias that was present in the earlier DLS photo-z calculations.  We provide the transmission curve electronically for use with other archival data which have used the $z$-band filter with the original MOSAIC cameras at KPNO and CTIO. 
We tested the improvement in photo-z performance using galaxies from DLS field F5, which overlaps the PRIMUS survey, and found a 20 per cent improvement in photo-z scatter at $\textrm{z}<0.6$ and a reduction in photo-z bias of more than a factor of two at $\textrm{z}>0.6$ (Figure~\ref{sigma_catoutlier}). Catastrophic outlier rates remained nearly unchanged.

One way of further reducing photo-z errors is by adding flux measurements in different passbands, beyond the original four DLS bands.  The addition of shorter wavelength bands ($u$-band, for example) that bracket the 4000$\textrm{\AA}$ break at low redshift can break degeneracies of the type described in Section~\ref{types}.  The near-infrared is also helpful in disambiguating galaxy types, so the addition of, e.g., $JHK_{S}$ data should also improve photo-z performance at low redshift, and the detection of the 4000$\textrm{\AA}$ break in the near-infrared can reduce the rate of catastrophic photo-z errors that result from misidentifying the Lyman break.  
We have observed $\simeq 8$ deg$^2$ of the DLS area in the $J$ and $K_s$ bands using the NEWFIRM detector on the CTIO and KPNO 4-m telescopes and the WIRC instrument on the Palomar 5-m.  We will examine the effect of including these bands in a future paper.  
One DLS field (F2) has been observed in the mid-IR using IRAC 4-band photometry, which will also be examined in the future; however, additional work to improve the mid-IR templates will be required in order for the photo-z's to be improved rather than degraded by the addition of IRAC imaging \citep[see, e.g.][]{badirac}.

Lacking spectroscopic redshifts in our training and test sets at $\textrm{z}>1$ we cannot evaluate photo-z performance in the DLS at high redshift.  However, there are several methods that can illuminate expected photo-z results.  We can compare our photo-z N(z) to deeper surveys such as COSMOS that have both deeper spectroscopic samples and a larger number of broad-band observations which produce higher precision photo-z's.  Using only bands similar to our $BVRz$ filter complement will enable us to estimate expected bias, scatter, and catastrophic outlier rates at fainter magnitudes.  Beyond direct comparison to representative spectroscopic samples, we can use spatial cross-correlations with a bright subset of galaxies \citep[e.g.,][Schmidt et al.~submitted]{Matt:10} to measure the redshift distribution of the survey.  However, the spectroscopic samples in question must span the redshift range of interest.  With both the SHELS and PRIMUS samples restricted to $\textrm{z} \lesssim 1$ we require further targeted spectroscopy in the DLS to fully utilize this technique.

As discussed earlier in the paper, much photometric redshift work to date has assumed that the template SEDs do not evolve and has used discretized types.  A few examples exist for fitting a continuous galaxy type \citep{Blan:07,Bram:08}, usually with linear combinations of a set of basis templates.  This form naturally allows for changes in the average template with redshift, but makes type dependent Bayesian priors more difficult to implement.  Methods of parametrizing template evolution and continuous galaxy type while simultaneously allowing for type-dependent priors will be explored in the future (Schmidt et al.~in prep.).  However, a larger number of narrower filter curves would be preferred for detailed examination of galaxy types and evolution.  Surveys such as J-PAS \citep[]{Ben:09} would be ideal for the study of template and prior evolution.

\section*{acknowledgements}
The authors would like to thank David Wittman, Tony Tyson, Dinesh Loomba, James Jee, Russell Ryan, Chris Morrison, Ami Choi, Vera Margoniner, Bego\~na Ascaso, and Jim Bosch for useful discussions and suggestions during the work discussed in this paper.

SJS was supported by NSF Grant AST-1009514.

PAT was partly supported by a grant from the New Mexico Space Grant Consortium.

Funding for the Deep Lens Survey has been provided
by Bell Labs Lucent Technologies and NSF grants AST 04-
41072 and AST 01-34753. Observations were obtained at
Cerro Tololo Inter-American Observatory and Kitt Peak National Observatory. 
CTIO and KPNO are divisions of the
National Optical Astronomy Observatory (NOAO), which
is operated by the Association of Universities for Research
in Astronomy, Inc., under cooperative agreement with the
National Science Foundation.

We would like to thank the SHELS team for making a portion of their spectroscopic sample available to us for use in the photometric redshift analysis.

We thank the PRIMUS team for sharing their redshift
catalogue. Funding for PRIMUS has been provided by NSF
grants AST-0607701, 0908246, 0908442, 0908354, and NASA
grant 08-ADP08-0019. This paper includes data gathered
with the 6.5-metre Magellan Telescopes located at Las Cam-
panas Observatory, Chile.

We thank CFHTLS for making their data available, observations obtained with MegaPrime/MegaCam, a joint project of CFHT and CEA/DAPNIA, at the Canada-France-Hawaii Telescope (CFHT) which is operated by the National Research Council (NRC) of Canada, the Institut National des Science de l'Univers of the Centre National de la Recherche Scientifique (CNRS) of France, and the University of Hawaii. This work is based in part on data products produced at TERAPIX and the Canadian Astronomy Data Centre as part of the Canada-France-Hawaii Telescope Legacy Survey, a collaborative project of NRC and CNRS.

\bibliography{draft.v2}

\begin{thebibliography}{43}
\expandafter\ifx\csname natexlab\endcsname\relax\def\natexlab#1{#1}\fi

\bibitem[{{Abrahamse} {et~al}\mbox{.}(2011){Abrahamse}, {Knox}, {Schmidt},
  {Thorman}, {Tyson}, \& {Zhan}}]{Abr:11}
{Abrahamse} A., {Knox} L., {Schmidt} S., {Thorman} P., {Tyson} J.~A., {Zhan}
  H., 2011, \apj, 734, 36

\bibitem[{{Aihara} {et~al}\mbox{.}(2011){Aihara}, {Allende Prieto}, {An},
  {Anderson}, {Aubourg}, {Balbinot}, {Beers}, {Berlind}, {Bickerton},
  {Bizyaev}, {Blanton}, {Bochanski}, {Bolton}, {Bovy}, {Brandt}, {Brinkmann},
  {Brown}, {Brownstein}, {Busca}, {Campbell}, {Carr}, {Chen}, {Chiappini},
  {Comparat}, {Connolly}, {Cortes}, {Croft}, {Cuesta}, {da Costa}, {Davenport},
  {Dawson}, {Dhital}, {Ealet}, {Ebelke}, {Edmondson}, {Eisenstein},
  {Escoffier}, {Esposito}, {Evans}, {Fan}, {Femen{\'{\i}}a Castell{\'a}},
  {Font-Ribera}, {Frinchaboy}, {Ge}, {Gillespie}, {Gilmore}, {Gonz{\'a}lez
  Hern{\'a}ndez}, {Gott}, {Gould}, {Grebel}, {Gunn}, {Hamilton}, {Harding},
  {Harris}, {Hawley}, {Hearty}, {Ho}, {Hogg}, {Holtzman}, {Honscheid}, {Inada},
  {Ivans}, {Jiang}, {Johnson}, {Jordan}, {Jordan}, {Kazin}, {Kirkby}, {Klaene},
  {Knapp}, {Kneib}, {Kochanek}, {Koesterke}, {Kollmeier}, {Kron}, {Lampeitl},
  {Lang}, {Le Goff}, {Lee}, {Lin}, {Long}, {Loomis}, {Lucatello}, {Lundgren},
  {Lupton}, {Ma}, {MacDonald}, {Mahadevan}, {Maia}, {Makler}, {Malanushenko},
  {Malanushenko}, {Mandelbaum}, {Maraston}, {Margala}, {Masters}, {McBride},
  {McGehee}, {McGreer}, {M{\'e}nard}, {Miralda-Escud{\'e}}, {Morrison},
  {Mullally}, {Muna}, {Munn}, {Murayama}, {Myers}, {Naugle}, {Fausti Neto},
  {Cuong Nguyen}, {Nichol}, {O'Connell}, {Ogando}, {Olmstead}, {Oravetz},
  {Padmanabhan}, {Palanque-Delabrouille}, {Pan}, {Pandey}, {P{\^a}ris},
  {Percival}, {Petitjean}, {Pfaffenberger}, {Pforr}, {Phleps}, {Pichon},
  {Pieri}, {Prada}, {Price-Whelan}, {Raddick}, {Ramos}, {Reyl{\'e}}, {Rich},
  {Richards}, {Rix}, {Robin}, {Rocha-Pinto}, {Rockosi}, {Roe}, {Rollinde},
  {Ross}, {Ross}, {Rossetto}, {S{\'a}nchez}, {Sayres}, {Schlegel},
  {Schlesinger}, {Schmidt}, {Schneider}, {Sheldon}, {Shu}, {Simmerer},
  {Simmons}, {Sivarani}, {Snedden}, {Sobeck}, {Steinmetz}, {Strauss}, {Szalay},
  {Tanaka}, {Thakar}, {Thomas}, {Tinker}, {Tofflemire}, {Tojeiro}, {Tremonti},
  {Vandenberg}, {Vargas Maga{\~n}a}, {Verde}, {Vogt}, {Wake}, {Wang}, {Weaver},
  {Weinberg}, {White}, {White}, {Yanny}, {Yasuda}, {Yeche}, \&
  {Zehavi}}]{aihara11}
{Aihara} H. {et~al.}, 2011, \apjs, 193, 29

\bibitem[{{Ben{\'{\i}}tez}(2000)}]{BPZ:00}
{Ben{\'{\i}}tez} N., 2000, \apj, 536, 571

\bibitem[{{Ben{\'{\i}}tez} {et~al}\mbox{.}(2009){Ben{\'{\i}}tez},
  {Gazta{\~n}aga}, {Miquel}, {Castander}, {Moles}, {Crocce},
  {Fern{\'a}ndez-Soto}, {Fosalba}, {Ballesteros}, {Campa}, {Cardiel-Sas},
  {Castilla}, {Crist{\'o}bal-Hornillos}, {Delfino}, {Fern{\'a}ndez},
  {Fern{\'a}ndez-Sopuerta}, {Garc{\'{\i}}a-Bellido}, {Lobo}, {Mart{\'{\i}}nez},
  {Ortiz}, {Pacheco}, {Paredes}, {Pons-Border{\'{\i}}a}, {S{\'a}nchez},
  {S{\'a}nchez}, {Varela}, \& {de Vicente}}]{Ben:09}
{Ben{\'{\i}}tez} N. {et~al.}, 2009, \apj, 691, 241

\bibitem[{{Blanton} \& {Roweis}(2007)}]{Blan:07}
{Blanton} M.~R., {Roweis} S., 2007, \aj, 133, 734

\bibitem[{{Bolzonella} {et~al}\mbox{.}(2000){Bolzonella}, {Miralles}, \&
  {Pell{\'o}}}]{Hyperz:00}
{Bolzonella} M., {Miralles} J.-M., {Pell{\'o}} R., 2000, \aap, 363, 476

\bibitem[{{Brammer} {et~al}\mbox{.}(2008){Brammer}, {van Dokkum}, \&
  {Coppi}}]{Bram:08}
{Brammer} G.~B., {van Dokkum} P.~G., {Coppi} P., 2008, \apj, 686, 1503

\bibitem[{{Bruzual} {et~al}\mbox{.}(retrieved 10/06/{2009}){Bruzual},
  {Persson}, {Gunn}, \& {Stryker}}]{bpgs}
{Bruzual}, {Persson}, {Gunn}, {Stryker}, retrieved 10/06/{2009}, {StSCI
  Calibration Database}

\bibitem[{{Budav{\'a}ri} {et~al}\mbox{.}(2000){Budav{\'a}ri}, {Szalay},
  {Connolly}, {Csabai}, \& {Dickinson}}]{Bud:00}
{Budav{\'a}ri} T., {Szalay} A.~S., {Connolly} A.~J., {Csabai} I., {Dickinson}
  M., 2000, \aj, 120, 1588

\bibitem[{{Choi} {et~al}\mbox{.}(2012){Choi}, {Tyson}, {Morrison}, {Jee},
  {Schmidt}, {Margoniner}, \& {Wittman}}]{Choi:12}
{Choi} A., {Tyson} J.~A., {Morrison} C.~B., {Jee} M.~J., {Schmidt} S.~J.,
  {Margoniner} V.~E., {Wittman} D.~M., 2012, \apj, 759, 101

\bibitem[{{Clampin} {et~al}\mbox{.}(1998){Clampin}, {Hartig}, {Ford},
  {Sirianni}, {Purdue}, {Walkowicz}, {Golimowski}, {Illingworth}, {Blouke},
  {Lesser}, {Burmester}, {Kimble}, {Sullivan}, \& {Krebs}}]{Clampin1998}
{Clampin} M. {et~al.}, 1998, in Society of Photo-Optical Instrumentation
  Engineers (SPIE) Conference Series, Vol. 3356, Society of Photo-Optical
  Instrumentation Engineers (SPIE) Conference Series, {P.~Y.~Bely \&
  J.~B.~Breckinridge}, ed., pp. 332--337

\bibitem[{{Coe} {et~al}\mbox{.}(2006){Coe}, {Ben{\'{\i}}tez}, {S{\'a}nchez},
  {Jee}, {Bouwens}, \& {Ford}}]{colorpro}
{Coe} D., {Ben{\'{\i}}tez} N., {S{\'a}nchez} S.~F., {Jee} M., {Bouwens} R.,
  {Ford} H., 2006, \aj, 132, 926

\bibitem[{{Coil} {et~al}\mbox{.}(2011){Coil}, {Blanton}, {Burles}, {Cool},
  {Eisenstein}, {Moustakas}, {Wong}, {Zhu}, {Aird}, {Bernstein}, {Bolton}, \&
  {Hogg}}]{coil:11}
{Coil} A.~L. {et~al.}, 2011, \apj, 741, 8

\bibitem[{{Coleman} {et~al}\mbox{.}(1980){Coleman}, {Wu}, \&
  {Weedman}}]{CWW:80}
{Coleman} G.~D., {Wu} C.-C., {Weedman} D.~W., 1980, \apjs, 43, 393

\bibitem[{{Covey} {et~al}\mbox{.}(2007){Covey}, {Ivezi{\'c}}, {Schlegel},
  {Finkbeiner}, {Padmanabhan}, {Lupton}, {Ag{\"u}eros}, {Bochanski}, {Hawley},
  {West}, {Seth}, {Kimball}, {Gogarten}, {Claire}, {Haggard}, {Kaib},
  {Schneider}, \& {Sesar}}]{covey07}
{Covey} K.~R. {et~al.}, 2007, \aj, 134, 2398

\bibitem[{{Fern{\'a}ndez-Soto} {et~al}\mbox{.}(2002){Fern{\'a}ndez-Soto},
  {Lanzetta}, {Chen}, {Levine}, \& {Yahata}}]{fern-soto02}
{Fern{\'a}ndez-Soto} A., {Lanzetta} K.~M., {Chen} H.-W., {Levine} B., {Yahata}
  N., 2002, \mnras, 330, 889

\bibitem[{{Geller} {et~al}\mbox{.}(2005){Geller}, {Dell'Antonio}, {Kurtz},
  {Ramella}, {Fabricant}, {Caldwell}, {Tyson}, \& {Wittman}}]{Gel:05}
{Geller} M.~J., {Dell'Antonio} I.~P., {Kurtz} M.~J., {Ramella} M., {Fabricant}
  D.~G., {Caldwell} N., {Tyson} J.~A., {Wittman} D., 2005, \apjl, 635, L125

\bibitem[{{Geller} {et~al}\mbox{.}(2010){Geller}, {Kurtz}, {Dell'Antonio},
  {Ramella}, \& {Fabricant}}]{Gel:10}
{Geller} M.~J., {Kurtz} M.~J., {Dell'Antonio} I.~P., {Ramella} M., {Fabricant}
  D.~G., 2010, \apj, 709, 832

\bibitem[{{Hildebrandt} {et~al}\mbox{.}(2010){Hildebrandt}, {Arnouts}, {Capak},
  {Moustakas}, {Wolf}, {Abdalla}, {Assef}, {Banerji}, {Ben{\'{\i}}tez},
  {Brammer}, {Budav{\'a}ri}, {Carliles}, {Coe}, {Dahlen}, {Feldmann}, {Gerdes},
  {Gillis}, {Ilbert}, {Kotulla}, {Lahav}, {Li}, {Miralles}, {Purger},
  {Schmidt}, \& {Singal}}]{badirac}
{Hildebrandt} H. {et~al.}, 2010, \aap, 523, A31

\bibitem[{{Hildebrandt} {et~al}\mbox{.}(2012){Hildebrandt}, {Erben}, {Kuijken},
  {van Waerbeke}, {Heymans}, {Coupon}, {Benjamin}, {Bonnett}, {Fu}, {Hoekstra},
  {Kitching}, {Mellier}, {Miller}, {Velander}, {Hudson}, {Rowe}, {Schrabback},
  {Semboloni}, \& {Ben{\'{\i}}tez}}]{hildebrandt12}
{Hildebrandt} H. {et~al.}, 2012, \mnras, 421, 2355

\bibitem[{{Hinkle} {et~al}\mbox{.}(2003){Hinkle}, {Wallace}, \&
  {Livingston}}]{atmo}
{Hinkle} K.~H., {Wallace} L., {Livingston} W., 2003, in Bulletin of the
  American Astronomical Society, Vol.~35, American Astronomical Society Meeting
  Abstracts, p. 1260

\bibitem[{{Ilbert} {et~al}\mbox{.}(2006){Ilbert}, {Arnouts}, {McCracken},
  {Bolzonella}, {Bertin}, {Le F{\`e}vre}, {Mellier}, {Zamorani}, {Pell{\`o}},
  {Iovino}, {Tresse}, {Le Brun}, {Bottini}, {Garilli}, {Maccagni}, {Picat},
  {Scaramella}, {Scodeggio}, {Vettolani}, {Zanichelli}, {Adami}, {Bardelli},
  {Cappi}, {Charlot}, {Ciliegi}, {Contini}, {Cucciati}, {Foucaud}, {Franzetti},
  {Gavignaud}, {Guzzo}, {Marano}, {Marinoni}, {Mazure}, {Meneux}, {Merighi},
  {Paltani}, {Pollo}, {Pozzetti}, {Radovich}, {Zucca}, {Bondi}, {Bongiorno},
  {Busarello}, {de La Torre}, {Gregorini}, {Lamareille}, {Mathez}, {Merluzzi},
  {Ripepi}, {Rizzo}, \& {Vergani}}]{Ilb:06}
{Ilbert} O. {et~al.}, 2006, \aap, 457, 841

\bibitem[{{Jee} {et~al}\mbox{.}(2012){Jee}, {Tyson}, {Schneider}, {Wittman},
  {Schmidt}, \& {Hilbert}}]{Jee:12}
{Jee} M.~J., {Tyson} J.~A., {Schneider} M.~D., {Wittman} D., {Schmidt} S.,
  {Hilbert} S., 2012, ArXiv e-prints

\bibitem[{{Kaiser} {et~al}\mbox{.}(2002){Kaiser}, {Aussel}, {Burke},
  {Boesgaard}, {Chambers}, {Chun}, {Heasley}, {Hodapp}, {Hunt}, {Jedicke},
  {Jewitt}, {Kudritzki}, {Luppino}, {Maberry}, {Magnier}, {Monet}, {Onaka},
  {Pickles}, {Rhoads}, {Simon}, {Szalay}, {Szapudi}, {Tholen}, {Tonry},
  {Waterson}, \& {Wick}}]{panstarrs}
{Kaiser} N. {et~al.}, 2002, in Society of Photo-Optical Instrumentation
  Engineers (SPIE) Conference Series, Vol. 4836, Society of Photo-Optical
  Instrumentation Engineers (SPIE) Conference Series, {J.~A.~Tyson \&
  S.~Wolff}, ed., pp. 154--164

\bibitem[{{Kinney} {et~al}\mbox{.}(1996){Kinney}, {Calzetti}, {Bohlin},
  {McQuade}, {Storchi-Bergmann}, \& {Schmitt}}]{Kin:96}
{Kinney} A.~L., {Calzetti} D., {Bohlin} R.~C., {McQuade} K., {Storchi-Bergmann}
  T., {Schmitt} H.~R., 1996, \apj, 467, 38

\bibitem[{{Laureijs} {et~al}\mbox{.}(2010){Laureijs}, {Duvet}, {Escudero Sanz},
  {Gondoin}, {Lumb}, {Oosterbroek}, \& {Saavedra Criado}}]{Lau:10}
{Laureijs} R.~J., {Duvet} L., {Escudero Sanz} I., {Gondoin} P., {Lumb} D.~H.,
  {Oosterbroek} T., {Saavedra Criado} G., 2010, in Society of Photo-Optical
  Instrumentation Engineers (SPIE) Conference Series, Vol. 7731, Society of
  Photo-Optical Instrumentation Engineers (SPIE) Conference Series

\bibitem[{{Le F{\`e}vre} {et~al}\mbox{.}(2005){Le F{\`e}vre}, {Vettolani},
  {Garilli}, {Tresse}, {Bottini}, {Le Brun}, {Maccagni}, {Picat}, {Scaramella},
  {Scodeggio}, {Zanichelli}, {Adami}, {Arnaboldi}, {Arnouts}, {Bardelli},
  {Bolzonella}, {Cappi}, {Charlot}, {Ciliegi}, {Contini}, {Foucaud},
  {Franzetti}, {Gavignaud}, {Guzzo}, {Ilbert}, {Iovino}, {McCracken}, {Marano},
  {Marinoni}, {Mathez}, {Mazure}, {Meneux}, {Merighi}, {Paltani}, {Pell{\`o}},
  {Pollo}, {Pozzetti}, {Radovich}, {Zamorani}, {Zucca}, {Bondi}, {Bongiorno},
  {Busarello}, {Lamareille}, {Mellier}, {Merluzzi}, {Ripepi}, \&
  {Rizzo}}]{vvds:05}
{Le F{\`e}vre} O. {et~al.}, 2005, \aap, 439, 845

\bibitem[{{Margoniner} \& {Wittman}(2008)}]{marg08}
{Margoniner} V.~E., {Wittman} D.~M., 2008, \apj, 679, 31

\bibitem[{{Matthews} \& {Newman}(2010)}]{Matt:10}
{Matthews} D.~J., {Newman} J.~A., 2010, \apj, 721, 456

\bibitem[{{McCracken} {et~al}\mbox{.}(2003){McCracken}, {Radovich}, {Bertin},
  {Mellier}, {Dantel-Fort}, {Le F{\`e}vre}, {Cuillandre}, {Gwyn}, {Foucaud}, \&
  {Zamorani}}]{McCvvds:03}
{McCracken} H.~J. {et~al.}, 2003, \aap, 410, 17

\bibitem[{{Morrison} {et~al}\mbox{.}(2012){Morrison}, {Scranton}, {M{\'e}nard},
  {Schmidt}, {Tyson}, {Ryan}, {Choi}, \& {Wittman}}]{Mor:12}
{Morrison} C.~B., {Scranton} R., {M{\'e}nard} B., {Schmidt} S.~J., {Tyson}
  J.~A., {Ryan} R., {Choi} A., {Wittman} D.~M., 2012, \mnras, 426, 2489

\bibitem[{{Padmanabhan} {et~al}\mbox{.}(2008){Padmanabhan}, {Schlegel},
  {Finkbeiner}, {Barentine}, {Blanton}, {Brewington}, {Gunn}, {Harvanek},
  {Hogg}, {Ivezi{\'c}}, {Johnston}, {Kent}, {Kleinman}, {Knapp}, {Krzesinski},
  {Long}, {Neilsen}, {Nitta}, {Loomis}, {Lupton}, {Roweis}, {Snedden},
  {Strauss}, \& {Tucker}}]{ubercal}
{Padmanabhan} N. {et~al.}, 2008, \apj, 674, 1217

\bibitem[{{Pickles}(1997)}]{pickles}
{Pickles} A.~J., 1997, VizieR Online Data Catalog, 7102, 0

\bibitem[{{Rajkanan} {et~al}\mbox{.}(1979){Rajkanan}, {Singh}, \&
  {Shewchun}}]{Rajkanan1979}
{Rajkanan} K., {Singh} R., {Shewchun} J., 1979, Solid State Electronics, 22,
  793

\bibitem[{{Ryan} {et~al}\mbox{.}(2011){Ryan}, {Thorman}, {Yan}, {Fan}, {Yan},
  {Mechtley}, {Hathi}, {Cohen}, {Windhorst}, {McCarthy}, \& {Wittman}}]{ryan11}
{Ryan} R.~E. {et~al.}, 2011, \apj, 739, 83

\bibitem[{{Scientific Imaging Technologies Inc.}(1995)}]{SiteCCDs}
{Scientific Imaging Technologies Inc.}, 1995, {Literature No. ST-002A}, 12th
  edn. {Retrieved from
  http://www.ociw.edu/instrumentation/ccd/parts/ST-002a.pdf}

\bibitem[{{Tyson}(2002)}]{lsst}
{Tyson} J.~A., 2002, in Society of Photo-Optical Instrumentation Engineers
  (SPIE) Conference Series, Vol. 4836, Society of Photo-Optical Instrumentation
  Engineers (SPIE) Conference Series, {J.~A.~Tyson \& S.~Wolff}, ed., pp.
  10--20

\bibitem[{{Varshni}(1967)}]{Varshni1967}
{Varshni} Y.~P., 1967, Physica, 34, 149

\bibitem[{{Wester} \& {Dark Energy Survey Collaboration}(2005)}]{DES}
{Wester} W., {Dark Energy Survey Collaboration}, 2005, in Astronomical Society
  of the Pacific Conference Series, Vol. 339, Observing Dark Energy, {Wolff}
  S.~C., {Lauer} T.~R., eds., p. 152

\bibitem[{{Williams} {et~al}\mbox{.}(1996){Williams}, {Blacker}, {Dickinson},
  {Dixon}, {Ferguson}, {Fruchter}, {Giavalisco}, {Gilliland}, {Heyer},
  {Katsanis}, {Levay}, {Lucas}, {McElroy}, {Petro}, {Postman}, {Adorf}, \&
  {Hook}}]{Will:96}
{Williams} R.~E. {et~al.}, 1996, \aj, 112, 1335

\bibitem[{{Wittman} {et~al}\mbox{.}(2007){Wittman}, {Riechers}, \&
  {Margoniner}}]{witt07}
{Wittman} D., {Riechers} P., {Margoniner} V.~E., 2007, \apjl, 671, L109

\bibitem[{{Wittman} {et~al}\mbox{.}(2012){Wittman}, {Ryan}, \&
  {Thorman}}]{wittman11}
{Wittman} D., {Ryan} R., {Thorman} P., 2012, \mnras, 421, 2251

\bibitem[{{Wittman} {et~al}\mbox{.}(2002){Wittman}, {Tyson}, {Dell'Antonio},
  {Becker}, {Margoniner}, {Cohen}, {Norman}, {Loomba}, {Squires}, {Wilson},
  {Stubbs}, {Hennawi}, {Spergel}, {Boeshaar}, {Clocchiatti}, {Hamuy},
  {Bernstein}, {Gonzalez}, {Guhathakurta}, {Hu}, {Seljak}, \&
  {Zaritsky}}]{wittman02}
{Wittman} D.~M. {et~al.}, 2002, in Society of Photo-Optical Instrumentation
  Engineers (SPIE) Conference Series, {J.~A.~Tyson \& S.~Wolff}, ed., Vol.
  4836, pp. 73--82

\end{thebibliography}
\bibliographystyle{mn2e}

\end{document}